\documentclass[aps,twocolumn,superscriptaddress]{revtex4-2}

\usepackage{amsmath,amssymb,color,graphicx,mathpazo,mathtools,times,float,dsfont,bbm,braket}
\usepackage[colorlinks=true,allcolors=blue]{hyperref}

\newcommand{\eref}[1]{Eq.~\eqref{eq:#1}}

\newcommand{\Fref}[1]{Figure~\ref{fig:#1}}
\newcommand{\fref}[1]{Fig.~\ref{fig:#1}}

\newcommand{\aref}[1]{App.~\ref{sec:#1}}

\begin{document}
\title{Quantum Zeno blockade in optomechanical systems}

\author{Karl Pelka}
\email{karl.a.pelka@um.edu.mt}
\affiliation{Department of Physics, University of Malta, Msida MSD 2080, Malta}
\author{Andr\'e Xuereb}
\email{andre.xuereb@um.edu.mt}
\affiliation{Department of Physics, University of Malta, Msida MSD 2080, Malta}
\date{\today}

\begin{abstract}
We investigate the application of the quantum Zeno effect (QZE) for the preparation of non-Gaussian states in optomechanical systems. By frequently monitoring the system, the QZE can suppress transitions away from desired subspaces of states. We show that this enables the preparation of states in qubit subspaces even in the presence of noise and decoherence. Through analytical and numerical analysis, we demonstrate that QZE-based protocols can significantly improve the robustness of state preparation of qubit states in continuous variable architectures. Our results extend the utility of the QZE beyond discrete systems, highlighting its potential for enhancing quantum control in more complex quantum information processing environments. These findings offer a promising approach for achieving reliable non-Gaussian states in optomechanical systems, with implications for the development of photonic quantum computing and quantum sensing.
\end{abstract}

\maketitle

\section{Introduction}
Originally conceived as a Gedankenexperiment showcasing the absurdity of frequent measurements in quantum mechanics~\cite{Misra1977}, the quantum Zeno effect (QZE) has shown in a multitude of systems to suppress the decay of unstable quantum systems~\cite{Itano1990,Fischer2001,Syassen2008,Yan2013,Barontini2013}, stop coherent dynamics~\cite{Streed2006,Raimond2012} or tunneling processes~\cite{Patil2015}. Another consequence is that the effect can inhibit the evolution of the system away from a desired Hilbert subspace, effectively "freezing" it in a particular set of states which allows the dynamical manipulation of the quantum system by frequently~\cite{Raimond2010,Signoles2014} or continuously~\cite{Bretheau2015} monitoring a quantum system~\cite{Facchi2008}. This aspect can be leveraged as a means of controling quantum transitions and manipulate quantum states with high precision for circuit QED systems~\cite{Raimond2012} and hybrid cavity-qubit-systems~\cite{Bretheau2015}. Among the various techniques available, the QZE has emerged as a powerful tool for robust state preparation in noisy environments which is fundamental to the realization of quantum technologies.  \par
Under the auspices of cavity optomechanics~\cite{Aspelmeyer2014}, recent theoretical and experimental advances have demonstrated the potential carried by systems experiencing radiation pressure for fundamental investigations as well as technological applications~\cite{Barzanjeh2022}. The radiation in a single mode of the electromagnetic radiation field, e.g., within a optical cavity with a high finesse~\cite{Kippenberg2005}, experiencing a dispersive shift and exerting a force on the motion of a harmonic oscillator through the radiation pressure force comprise the prototypical optomechanical system~\cite{Marquardt2006}. The mechanical element comes in many guises such as a micro- or nanoparticle in the cavity~\cite{Millen2014,Delic2019}, a semi-transparent membrane inside the cavity~\cite{Thompson2008}, one of the end mirrors of the cavity~\cite{Groeblacher2009}, one plate of a capacitor~\cite{Teufel2011}, or the cavity itself in the case of micro-toroids supporting whispering gallery modes of the radiation field~\cite{Kippenberg2005}. The combination of quantum-noise-limited optical control with high mechanical coherence allows for quantum transducers~\cite{Bochmann2013,Andrews2014}, directional amplifiers for microwave radiation~\cite{Barzanjeh2017,Bernier2017,Malz2018,Mercier2019}, and enables mechanical frequency combs~\cite{Allain2021}. Optomechanical interactions can be exploited to create entanglement between macroscopic mechanical oscillators~\cite{Ockeloen2018,Riedinger2018,Orr2023,Pelka2024}, cool mechanical systems to their quantum ground state~\cite{Teufel2011,Chan2011} and create parametric oscillations~\cite{Schliesser2006}. Multimode systems can exploit these self-sustained oscillations for synchronisation phenomena~\cite{Heinrich2011,Zhang2012,Zhang2015,Loerch2017,Pelka2020} and can display distinct behaviour based on temporal driving schemes~\cite{Malz2016,Xu2019,Mercade2021,Pelka2022} or many-body dynamics in complex networks~\cite{Schmidt2015,Peano2015,Mathew2020}. Radiation pressure can also be used to realise a variety of interactions between the constituents beyond linear dispersive shifts such as quadratic ones~\cite{Thompson2008,Sankey2010,Wilson2009,Reinhardt2016,Purdy2010} or cross-Kerr couplings~\cite{Heikkila2014,Solki2023}.\par
In this paper, we study the restriction of bosonic systems to a subset of possible states through nonlinearities. We show that this effectively achieves a phonon blockade, distinct from the photon blockade in the strong optomechanical coupling regime~\cite{Rabl2011}. We show that the required parameter regime can be achieved in state-of-the-art experiments and provide a technique for potential applications of optomechanical systems as quantum nonlinear devices, with significant relevance for optical quantum computation, photonic quantum simulation schemes, or quantum information processing~\cite{Stannigel2012}. \par
We proceed by discussing the minimal setup to generate the quantum Zeno effect in bosonic systems through strong continuous couplings~\cite{Facchi2003}. This analysis will allow us to gain the intuition for the occuring physics. We will then show that the effect carries over to achievable optomechanical couplings schemes that will introduce the effective phonon blockade. We then show through numerical analysis, that the blockade can be tested in state-of-the-art experiments in spite of operating in an open environment and far from the optomechanical strong coupling regime. Ultimately, we conclude with a discussion on the implication of our results.
\begin{figure}
\begin{center}
\includegraphics[width=0.99\columnwidth]{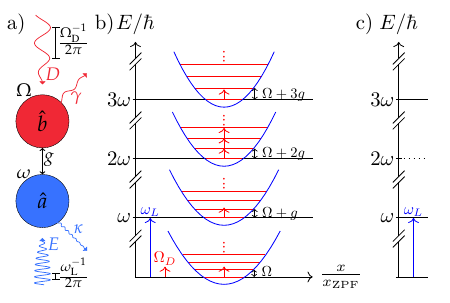}
\end{center}
\caption{(Color online) Quantum Zeno effect in an optomechanical cavity. (a) An optomechanical system consisting of a mechanical and an optical mode both driven at distinct frequencies coupled nonlinearly. (b) A cross-Kerr nonlinearity enables a mechanical drive to address mechanical transitions solely for a particular photon number. (c) The resonant states of the mechanical drive form a Zeno subspace which makes them inaccessible from the rest of the states.}
\label{fig:Model}
\end{figure}

\section{Effective model}
\begin{figure*}
\centering
\includegraphics[width=0.99\textwidth]{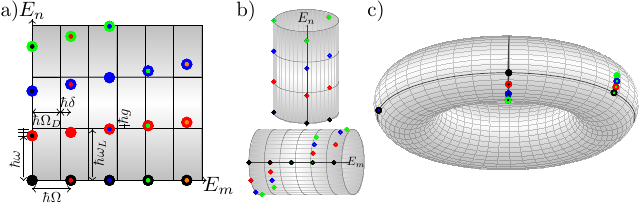}
\caption{(Color online) Emergence of Quantum Zeno Dynamics through cross-Kerr interaction. a) Two-dimensional arrangement of energy eigenvalues show that the energy spacing for the addition of a photon grows with the phonon number. b) The invariant Zeno subspaces are found by coalescence of the energy eigenvalues when rolling up the energy plane with the periodicity given by the drive frequencies. c) We see that the laser drive allows to select the invariant subspace, here all states with two phonons indicated by a blue interior, which are inaccessible from the remaining levels.}
\label{fig:ZenoBlockade}
\end{figure*}

We consider the dynamics of two driven bosonic modes that are described by the Hamiltonian

\begin{equation}
\frac{\hat{H}}{\hbar}=\omega\hat{a}^{\dagger}\hat{a}+\Omega\hat{b}^{\dagger}\hat{b}+g\hat{O}+[iEe^{i\omega_Lt} \hat{a}+iDe^{i\Omega_Dt} \hat{b}+\text{h.c.}]
\label{eq:Hamiltonian}
\end{equation}

with $\hat{a}$ ($\hat{b}$) describing the bosonic annihilation operators of the optical (mechanical) mode, $\omega$ ($\Omega$) their respective resonance frequencies, $g$ the strength of the coupling described by the operator $\hat{O}$. The parameters $|E|=\sqrt{2\mathcal{P}_L\kappa/\hbar\omega_L}$ ($|D|=\sqrt{2\mathcal{P}_D\gamma/\hbar\Omega_D}$) describe strength of the optical (mechanical) drive occurring at $\omega_L$ ($\Omega_D$) in relation to the drive's input power $\mathcal{P}_L$ ($\mathcal{P}_D$) and mode's decay rate $\kappa$ ($\gamma$). We proceed to analyze our system using the strong continuous coupling mechanism that leads to the quantum Zeno effect~\cite{Facchi2003}. This approach relies on the partition of a Hamiltonian into a measurement Hamiltonian $\hat{H}_{c}$ that is responsible for the continuous measurement while the remainder $\hat{H}_o(t)$ constitutes the system under observation. In the limit of strong coupling, the evolution operator $\mathcal{U}_K(t)=\mathcal{T}\exp(-\frac{i}{\hbar}\int_{0}^{t}(K\hat{H}_{c}+\hat{H}_{o}(\tau))\text{d}\tau)$ approaches the limiting operator $\mathcal{U}_Z(t)\lim_{K\rightarrow\infty}\mathcal{U}_K(t)=\mathcal{T}\exp(-\frac{i}{\hbar}\int_{0}^{t}\hat{H}_Z(\tau) \text{d}\tau)$ with the Zeno Hamiltonian
\begin{align}
\hat{H}_Z(t)=K\hat{H}_{c}+\sum_j\hat{P}_j\hat{H}_{o}(t)\hat{P}_j
\label{eq:ZenoHam}
\end{align}
where $\hat{P}_j$ denotes the projection operators onto the eigenspaces of $\hat{H}_{c}$ with the respective eigenvalue $\eta_{j}$. Since this projection turns the Hamiltonian of the observed system into block diagonal form there can be no transition between the distinct eigenspaces $\mathcal{H}_{\hat{P}_j}$ and the Zeno effect occurs. \par
We counterintuitively identify the system under observation with the drive $\hat{H}_{o}(t)=[i\hbar Ee^{i\omega_Lt} \hat{a}+i\hbar De^{i\Omega_Dt} \hat{b}+\text{h.c.}]$ and the measurement Hamiltonian with the bosonic modes and their interaction $\hat{H}_{c}=\hbar\omega\hat{a}^{\dagger}\hat{a}+\hbar\Omega\hat{b}^{\dagger}\hat{b}+\hbar g\hat{O}$ to conduct the analysis of the system described by \eref{Hamiltonian}. We begin the investigation with a cross-Kerr interaction described by $\hat{O}=\hat{a}^{\dagger}\hat{a}\hat{b}^{\dagger}\hat{b}$ as displayed in \fref{Model}. In this case, number states $\ket{n}\otimes\ket{m}$ remain eigenstates of $\hat{H}_{c}$ and the energy spectrum is $E_{nm}=\hbar(\omega n+ \Omega m + gmn)$ as displayed in \fref{Model}(b). This interaction allows us to go into the rotating frame, as outlined in \aref{AppA}, and rewrite the Zeno Hamiltonian in \eref{ZenoHam} in a suggestive manner

\begin{align}
\hat{H}_B/\hbar=\Delta\hat{a}^{\dagger}\hat{a}+(\delta+g\hat{a}^{\dagger}\hat{a})\hat{b}^{\dagger}\hat{b}+\sum_{k}\hat{P}_{k}\hat{H}_{o}\hat{P}_{k}.
\label{eq:ZenoBlock}
\end{align}

with the optical (mechanical) detuning $\Delta=\omega-\omega_L$ ($\delta=\Omega-\Omega_D$). The strong coupling limit requires in this case that the driving strengths $D$ and $E$ have to be much smaller than the frequencies $\omega$, $\Omega$, and $g$.

We can determine the frequency spectrum of this Hamiltonian as $\tilde{\omega}_{nm}=\Delta n+(\Omega-\Omega_D+gn) m$ and see that the frequency of the mechanical drive can be tuned to enable a degeneracy: If the drive frequency is tuned to $\Omega_D=\Omega+gN$ with $N\in\mathbb{N}$ we find that all states $\ket{N}\otimes\ket{m}$ are degenerate with the eigenvalue $\eta_{N}=\Delta N$. It can be seen in \fref{Model}(b) that the mechanical drive with this particular frequency is resonant with the energy ladder of the mechanical oscillator if and only if $N$ photons are present in the optical mode. Moreover, we find that for any finite optical detuning ($\Delta\ne0$) the eigenvalue is different from all states with $n\ne N$ as 
\begin{align}
\tilde{\omega}_{nm}=\Delta n+(g(n-N))m.
\end{align}
This means that \textit{the mechanical drive enables the tunable removal of the state with $N$ excitations from the optical mode for} $\Omega_D=\Omega+gN$ according to the formation of the quantum Zeno effect from strong continuous coupling which is illustrated in \fref{Model}(c). \par
To give this result a more physical interpretation we evaluate the Hamiltonian in this removed subspace, described by the projector $\hat{P}_{Nm}=\ket{N}\bra{N}\otimes\mathbb{1}_B$. We find in this case that $\hat{H}_N/\hbar = \Delta \hat{a}^{\dagger}\hat{a} +iD(\hat{b}^{\dagger}-\hat{b})$ and see that the mechanical system would be driven if the optical state acquired the removed amount of excitations. Thus, the mechanism behind this quantum Zeno effect is revealed to lie in the resonance of the mechanical drive: if the optical system were to achieve the amount of excitations such that the mechanical drive becomes resonant with the mechanical mode, the drive would transmit energy into the mechanical mode. Since this transmission of energy is only possible on resonance it would enable to draw conclusions about the optical state constituting a measurement. \par
It is worth noticing that the cross-Kerr interaction maintains the symmetry of the measurement Hamiltonian under exchange of the operators $\hat{a}\leftrightarrow\hat{b}$ and therefore the \textit{optical drive likewise enables the tunable removal of the state with $M$ excitations from the mechanical mode for} $\omega_L=\omega+gM$. The structure of the distinct Zeno subspaces can be determined geometrically as discussed in \aref{AppA} and is displayed for this case in \fref{ZenoBlockade}. The energy eigenvalues $E_{nm}$ are arranged on a two dimensional plane in \fref{ZenoBlockade}(a) such that the sum of the cartesian coordinates add up to the respective value. Here, the color of the inner circle describes the optical quantum number $n$ and the outer ring has the same color for all states with the same mechanical quantum number $m$. \fref{ZenoBlockade}(b) illustrates the plane wrapped up with the periodicities given by the drive frequency $\Omega_D$ in the upper diagram and with $\omega_L=\omega+2g$ in the lower diagram. Note that in the upper case all points are scattered around the cylinder while in the lower case all points all eigenvalues with the mechanical quantum number $M=2$, indicated by a blue interior, coalesce. Wrapping up each cylinder around the remaining axis results in the torus shown in \fref{ZenoBlockade}(c). The invariant Zeno subspaces can be determined by the coalescence of the respective eigenvalues, in this case $span(\ket{n}\otimes\ket{2})$, and those states become inaccessible from the remaining states in the strong coupling limit.

\section{Optomechanical realization}
To confirm that the quantum Zeno effect remains with finite coupling strengths and decay into a thermal environment, we conduct numerical simulations of the corresponding quantum optical master equation. The optical (mechanical) mode decays with the rate $\kappa$ ($\gamma$) into its bath at temperature $T$, realized with the standard dissipator in Lindblad form $\mathcal{D}_{\hat{o}}[\rho]=\hat{o}\rho\hat{o}^{\dagger}-\frac{1}{2}\{\rho,\hat{o}^{\dagger}\hat{o}\}$. The dynamics of the systems density matrix $\rho$ are described by the master equation in Lindblad form

\begin{multline}
\dot{\rho}=-\frac{i}{\hbar}[\hat{H},\rho] + \kappa\{(\bar{N}+1)\mathcal{D}_{\hat{a}}[\rho]+\bar{N}\mathcal{D}_{\hat{a}^{\dagger}}[\rho]\}\\
+\gamma\{(\bar{M}+1)\mathcal{D}_{\hat{b}}[\rho]+\bar{M}\mathcal{D}_{\hat{b}^{\dagger}}[\rho]\}.
\label{MasterEQN}
\end{multline}

where the temperature $T$ determines the average occupation $\bar{N}=[\exp(\hbar\omega/k_BT)-1]^{-1}$ ($\bar{M}=[\exp(\hbar\Omega/k_BT)-1]^{-1}$) of the optical (mechanical) mode. Throughout the simulations, we employ parameters adapted from~\cite{Heikkila2014,Pirkkalainen2015,Solki2023} and the quantum ground state $\rho_0=\ket{0}\bra{0}\otimes\ket{0}\bra{0}$ is used as the initial state. Concretely, we consider a cavity with an optical frequency of $\omega/2\pi=5$ GHz and decay rate $\kappa/2\pi=64.8$ kHz and a mechanical mode with frequency $\Omega/2\pi=65$ MHz and decay rate $\gamma/2\pi=10$ kHz coupled with and the couplings strength $g/2\pi=2.7$ MHz to each other. Both modes are in contact with environments thermalized at $T=20$ mK which amounts to $\bar{M}=0.267$ and $\bar{N}=6.46\times10^{-6}$. Further nonlinearities that could occur and their negligible impact on the scheme are discussed in \aref{AppB}. 
\begin{figure}
\centering
\includegraphics[width=0.485\textwidth]{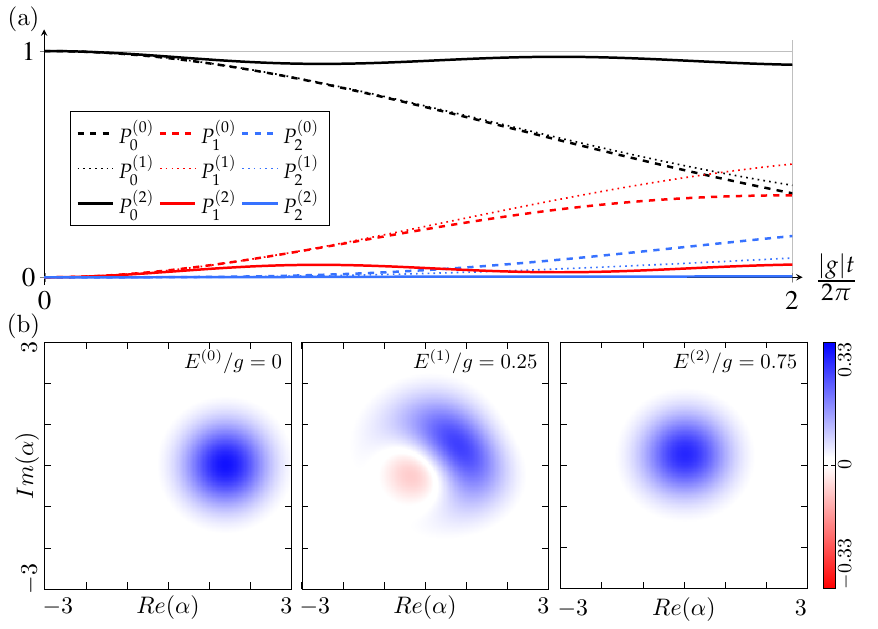}
\caption{(Color online) Analysis of resulting state from the quantum Zeno dynamics for the open system driven with $D^{(2)}/g=0.065$ at frequency $\Omega_{D}=\Omega$. a) Probability to find zero (black), one (red), and two (blue) phonons in the mechanical subsystem over time for optical drive with $E^{(0)}/g=0$ (dashed), $E^{(1)}/g=0.25$ (dotted), and $E^{(2)}/g=0.75$ (solid) at frequency $\omega_{L}=\omega+2g$ and. b) Resulting Wigner function for $E^{(0)}/g=0$ in the left panel showing a displacement of the mechanical state, $E^{(1)}/g=0.25$ in the middle panel showing the occurrence of negative patches indicating the non-Gaussian state, and $E^{(2)}/g=0.75$ disabling the displacement of the mechanical state.}
\label{fig:ZenoDyn}
\end{figure}
\Fref{ZenoDyn} summarizes the characteristics of the resulting state of the mechanical oscillator $\rho_B(t)=\text{Tr}_{A}\rho(t)$ for the mechanical drive $D^{(1)}/g=0.02$ 
at the frequency $\Omega_D=\Omega$ for optical drives of varying strength $E^{(0)}/g=0$ (dashed), $E^{(1)}/g=0.25$ (dotted), and $E^{(2)}/g=0.75$ (solid) at the frequency of the second photon level $\omega_L=\omega+2g$. \fref{ZenoDyn}(a) shows the occupation probabilities of the mechanical ground state $P_0$ (black), the mechanical state containing one phonon $P_1$ (red), and the second excited mechanical state $P_2$ (blue). For absent optical drive (dashed), the mechanical drive excites phonons leading to an increase in the probability of one and two phonons, while the probability to remain in the ground state deteriorates. The left panel of \fref{ZenoDyn}(b) shows the corresponding Wigner function and that without an optical drive establishing a phonon blockade the final state of the mechanical subsystem is a displaced state. Establishing the blocking tone at the intermediate drive amplitude $E^{(1)}$ (dotted lines), we see the increase to find one or zero phonons in the subsystem while state with two phonons is suppressed while more than two phonons do not occur. This signifies that \textit{the phonon blockade due to the quantum Zeno effect through intermediate optical drive at $\omega_{L}=\omega+Mg$ makes states with more than $M$ phonons effectively inaccessible} from the quantum ground state $\rho_0$. The middle panel of \fref{ZenoDyn}(b) shows that the corresponding Wigner function has a negative patch signifying the quantum nature of the state in the mechanical subsystem. Further increasing the optical drive amplitude to $E^{(2)}$ (solid lines) results in a strong suppression of the first excited phonon state and a complete suppression of the second phonon state. The right panel of \fref{ZenoDyn}(b) shows that with the increased power of the blocking tone, the final state has negligible occupation for more than one phonon and the quantum Zeno blockade is in full effect. \par
The observation that the phonon blockade for intermediate couplings effectively suppresses the occupation of more than $M$ phonons in the mechanical subsystem can be used to restrict the mechanical subsystem to the ground state and first excited state, effectively modeling a qubit. Thus, we simulate the system with the optical drive amplitude $E^{(2)}/g=0.75$ at frequency $\omega_{L}=\omega+g$ and mechanical drive $D^{(2)}/g=0.065$ on resonance ($\Omega_D=\Omega$) while all other parameters remain the same and find the behaviour illustrated in \fref{ZenoState}. In \fref{ZenoState}(a), we see that the driving conditions lead to a state where probability for more than one phonon in the cavity is strongly suppressed while the probability for one phonon is dominant. The resulting Wigner function is shown in \fref{ZenoState}(b) and shows a negative patch in the center of phase space, reminiscent of the Wigner function of the Fock state $\ket{1}$. Indeed, we show in \aref{AppC} that a cavity with decreased linewidth $\tilde{\kappa}=0.1\kappa$ allows the creation of the Fock state $\ket{1}$ with a fidelity of larger than 0.9 by using an optical drive with multiple frequencies and intermediate amplitudes. Our investigations show that the quantum Zeno blockade allows to create genuine mechanical quantum states required in quantum sensing with current state-of-the-art experiments while rapid advances in experimental designs bear the potential to controllably create mechanical Fock states with high fidelity.

\begin{figure}
\centering
\includegraphics[width=0.455\textwidth]{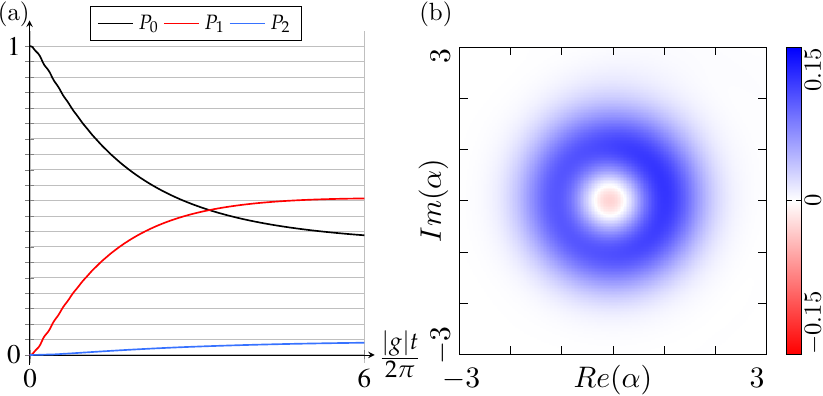}
\caption{(Color online) Analysis of resulting state from the quantum Zeno dynamics for the open system driven at $E^{(2)}/g=0.75$ at frequency $\omega_{L}=\omega+g$ and $D^{(2)}/g=0.065$ at frequency $\Omega_{D}=\Omega$. a) Probability to find zero (black), one (red), and two (blue) phonons in the mechanical subsystem over time showing a strong suppression of two or more phonons. b) Resulting Wigner function of the final state displaying a clear negative patch signalling the non-Gaussian nature of the state in the mechanical subsystem.}
\label{fig:ZenoState}
\end{figure}

\section{Conclusion and outlook}
We have demonstrated the quantum Zeno effect in an optomechanical system with the goal of preparing qubit states within a purely bosonic system. Our results suggest that quantum Zeno dynamics provide a viable pathway for precise qubit initialization by leveraging the inherent quantum dynamics even in the presence of environmental noise and decoherence. The findings open up new avenues for integrating QZE-based protocols in state of the art optomechanical systems, enabling them to be used in qubit-based quantum information processing offering broad potential across various quantum computing platforms. Optomechanical systems with the standard dispersive shift may benefit from the Zeno blockade by easing the requirements to obtain a photon blockade while systems realizing the quadratic coupling regime offer further promising platforms~\cite{Thompson2008,Sankey2010,Wilson2009,Reinhardt2016,Purdy2010} to realise the proposed scheme and remain to be explored. As cross-Kerr couplings are common in Josephson junction coupled microwave cavities~\cite{Holland2015}, our proposed scheme could enhance their capabilities as well. All platforms that realize our proposed scheme can readily test the potential to stabilize the ground state for increasing bath temperature and the possibility to generate Fock states by successive shaping of the available Zeno subspaces. We aim to pursue all of these avenues in future works, as well as the correspondence of the proposed blockade enabling only single excitations with genuine fermionic systems.

\begin{acknowledgments}
KP and AX would like to thank Peter Rabl and Shabir Barzanjeh for helpful discussions and acknowledge financial support from the Ministry for Education, Sports, Youths, Research and Innovation of the Government of Malta through its participation in the QuantERA ERA-NET Cofund in Quantum Technologies (project MQSens) implemented within the European Union’s Horizon 2020 Programme.
\end{acknowledgments}

\appendix

\section{Formation of Zeno subspaces in bosonic systems}
\label{sec:AppA}

\begin{figure*}
 \centering
 \includegraphics[width=0.86\linewidth]{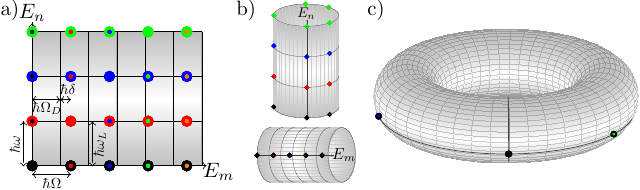}
 \caption{(Color online) Visualisation of the formation of invariant quantum Zeno subspaces for driven bosonic modes. a) The energy eigenvalues for two bosonic modes can be arranged in a two dimensional plane and the energy can be determined through the sum of the two coordinates. b) The behaviour of the drive Hamiltonian continuously observed through the bosonic modes can be determined by rolling up the energy eigenvalue plane with respect to the frequencies of both drives. If the drive frequency does not match the frequency spacing as shown in the upper diagram, the energy eigenvalues sweep the cylinder and hence the drive does not allow transitions among the corresponding states. If the energy eigenvalues coalesce as in the lower case, the states can be accessed through the drive. c) The drive generates invariant quantum Zeno subspaces and the dynamics within each subspace are given by the corresponding Zeno Hamiltonian.}
\label{fig:QuantumZeno}
\end{figure*}

In order to understand the formation of the Zeno subspaces that evolve independent of each other, we discuss the consequences of a strong coupling mechanism~\cite{Facchi2003} for the bosonic modes. We note that a time dependent change of basis commutes with the determination of the Zeno Hamiltonian in \eref{ZenoHam}, if the unitary $U(t)$ that describes the basis change commutes with all projectors $[U(t),P_j]=0$. To apply this theorem to our system, we assume that the coupling operator $\hat{O}$ commutes with the projector onto any number state $P_{nm}=(\ket{n}\otimes\ket{m}) (\bra{n}\otimes\bra{m})$ which allows us to go into the frame rotating with the optical frequency $\omega_L$ and mechanical frequency $\Omega_D$. This is described by the unitary transformation $U(t)=\exp(-i(\omega_L\hat{a}^{\dagger}\hat{a}+\Omega_D\hat{b}^\dagger\hat{b})t)$ and we find the Zeno Hamiltonian

\begin{align}
\hat{H}_Z/\hbar=\Delta\hat{a}^{\dagger}\hat{a}+\delta\hat{b}^{\dagger}\hat{b}+g\hat{O}+\sum_{k}\hat{P}_{k}\hat{H}_{o}\hat{P}_{k},
\label{eq:ZenoHamiltonian}
\end{align}

The projected Hamiltonian of the system under observation now depends on the spectrum of the measurement Hamiltonian. In absence of any coupling $\hat{O}$, the frequency spectrum is given in terms of the quantum numbers $n,m$ as $\omega_{n,m}=\Delta n+\delta m$ and we find that the detuning parameters determine the occuring degeneracies. In the case that both detunings vanish, the spectrum is fully degenerate and the single projector $\hat{P}_{1}$ onto the eigenvalue $0$ is the identity operator on the entire Hilbert space $\mathbb{1}_\mathcal{H}$ which leads to the Zeno Hamiltonian
\begin{align}
\hat{H}_Z=\hat{P}_{1}\hat{H}_{o}\hat{P}_{1}=\hat{H}_{o}=i\hbar E(\hat{a}^{\dagger}-\hat{a})+i\hbar D(\hat{b}^{\dagger}-\hat{b}).
\end{align}
This results in the Zeno evolution $\mathcal{U}_Z(t)=\hat{D}_{a}(iEt)\hat{D}_{b}(iDt)$ with the displacement operators $\hat{D}_{a}(\alpha)=\exp(\alpha^*\hat{a}-\alpha\hat{a}^{\dagger})$ and $\hat{D}_{b}(\beta)=\exp(\beta^*\hat{b}-\beta\hat{b}^{\dagger})$. In the case that only one detuning is vanishing, say $\Delta=0$, we find that there exist countably many subspaces enumerated with the eigenvalue $\eta_m=\delta m$ with the corresponding projector $P_m=\mathbb{1}_{a}\otimes (\ket{m} \bra{m})$. We find the Zeno Hamiltonian $\hat{H}_Z/\hbar=\delta\hat{b}^{\dagger}\hat{b}+iE(\hat{a}^{\dagger}-\hat{a})$ which shows the partition of the Hilbert space $\mathcal{H}$ into subspaces $\mathcal{H}_m$ that are evolving independently. The evolution of any pure state $\ket{\Psi(0)}=\ket{\psi_a(0)}\otimes(\sum_mc_m\ket{m})$ for instance is given by $\ket{\Psi(t)}=\hat{D}_{a}(iEt)\ket{\psi_a(0)}\otimes(\sum_mc_me^{im\delta t}\ket{m})$. Likewise, the case where the mechanical detuning is vanishing leads to a Hamiltonian that evolves independently in the subspaces with a given photon number and acts as a displacement operator $\hat{D}_{b}(iDt)$ in each of the subspaces. If neither detuning is vanishing (and they are not multiples of each other), all eigenvalues are distinct and the projection takes place over all projectors $\hat{P}_{nm}=(\ket{n}\otimes\ket{m}) (\bra{n}\otimes\bra{m})$ which results in the vanishing of the dynamics $\sum_{n,m}\hat{P}_{nm}\hat{H}_{o}\hat{P}_{nm}=0_{\mathcal{H}}$ of the system under observation. This analysis recovers the physical intuition behind the drive Hamiltonian $\hat{H}_{o}(t)$ that it acts as a displacement of the initial state with an amplitude increasing in time if and only if the respective drive is resonant. We find that the formation of the invariant subspaces can be understood with a geometrical construction depicted in \fref{QuantumZeno}: Instead of arranging the energy eigenvalues onto a one dimensional line $(E_{nm})$, it is possible to arrange the eigenvalues onto a two-dimensional plane $(E_n,E_m)$ such that the total energy is given as the sum of the cartesian coordinates $E_{nm}=E_n+E_m$ as shown in \fref{QuantumZeno}(a). The behaviour of the drive Hamiltonian $\hat{H}_o$ continuously observed through the bosonic modes  $\hat{H}_c$ is determined by wrapping up the energy eigenvalue plane periodically with respect to the frequencies of both drives $\omega_L$ and $\Omega_D$ which is depicted in \fref{QuantumZeno}(b). As the drive frequency $\Omega_D$ does not match the frequency spacing $\Omega$ as shown in the upper diagram, the energy eigenvalues sweep the cylinder and hence the drive does not allow transitions among the corresponding states. States can only be accessed through the drive if the energy eigenvalues coalesce through this wrapping process as shown in the lower diagram. The complete analysis of the invariant quantum Zeno subspaces is shown in \fref{QuantumZeno}(c) and requires to wrap up the plane along both directions which results in the energy eigenvalues distributed on a torus. Invariant Zeno subspaces are determined by the coalescence of the energy eigenvalues and the dynamics within each subspace are given by the corresponding Zeno Hamiltonian.

\section{Analysis of additional perturbing interactions}
\label{sec:AppB}
As our optomechanical realisation is based on the experiments discussed in~\cite{Heikkila2014}, it is worth investigating that it may come with additional nonlinear interactions that would lead to a Hamiltonian as outlined in~\cite{Solki2023}

\begin{equation}
\frac{\hat{H}_{P}}{\hbar}=\omega\hat{a}^{\dagger}\hat{a}+\Omega\hat{b}^{\dagger}\hat{b}+g\hat{a}^{\dagger}\hat{a}\hat{b}^{\dagger}\hat{b}+\chi \hat{a}^{\dagger}\hat{a}(\hat{b}^{\dagger}\hat{b})^2.
\label{eq:HamiltonianPRL}
\end{equation}

\begin{figure}
\centering
\includegraphics[width=0.485\textwidth]{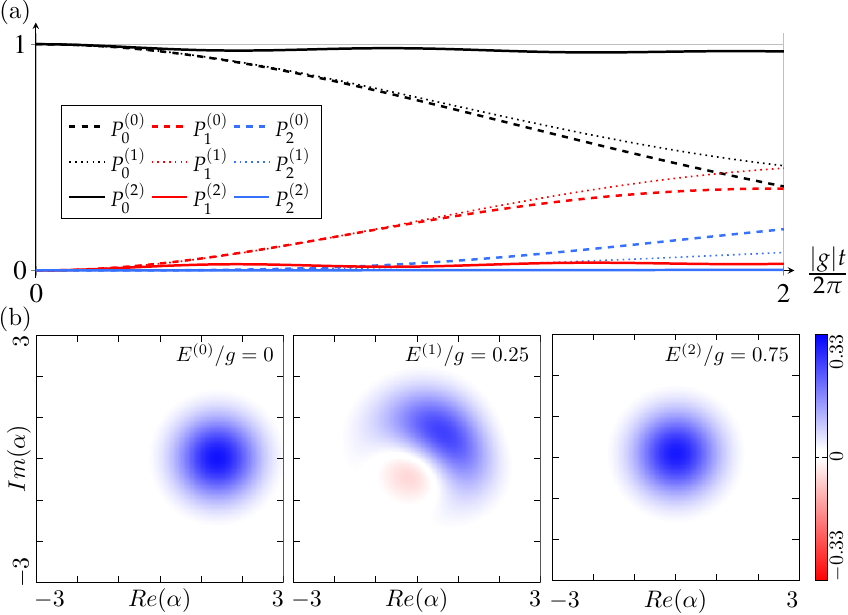}
\caption{(Color online) Analysis of resulting state from the quantum Zeno dynamics for the open system driven with $D^{(2)}/g=0.065$ at frequency $\Omega_{D}=\Omega$. a) Probability to find zero (black), one (red), and two (blue) phonons in the mechanical subsystem over time for optical drive with $E^{(0)}/g=0$ (dashed), $E^{(1)}/g=0.25$ (dotted), and $E^{(2)}/g=0.75$ (solid) at frequency $\omega_{L}=\omega+2g+4\chi$ and. b) Resulting Wigner function for $E^{(0)}/g=0$ in the left panel showing a displacment of the mechanical state, $E^{(1)}/g=0.25$ in the middle panel showing the occurrence of negative patches indicating the non-Gaussian state, and $E^{(2)}/g=0.75$ disabling the displacement of the mechanical state.}
\label{fig:ZenoDynEXTRA}
\end{figure}

We conduct the analysis of the quantum Zeno effect by designating $\hat{H}_{P}$ as the measurement Hamiltonian $\hat{H}_{c}$ and the system under observation is described by $\hat{H}_{o}(t)$. The resulting spectrum derived in the frame rotating with the drive frequencies $\omega_L$ and $\Omega_D$ is given by

\begin{align}
\tilde{\omega}_{nm}=(\Omega-\Omega_D) m+(\omega-\omega_L+ g m+\chi m^2)n.
\end{align}

This spectrum shows that an optical drive tuned to $\omega_L=\omega+Mg+M^2\chi$ makes all states with $M$ phonons degenerate. Hence, we find that in this optomechanical system the optical drive at frequency $\omega_L=\omega+Mg+M^2\chi$ removes the state with $M$ phonons through the quantum Zeno effect from all remaining states. \par

\begin{figure}
\centering
\includegraphics[width=0.485\textwidth]{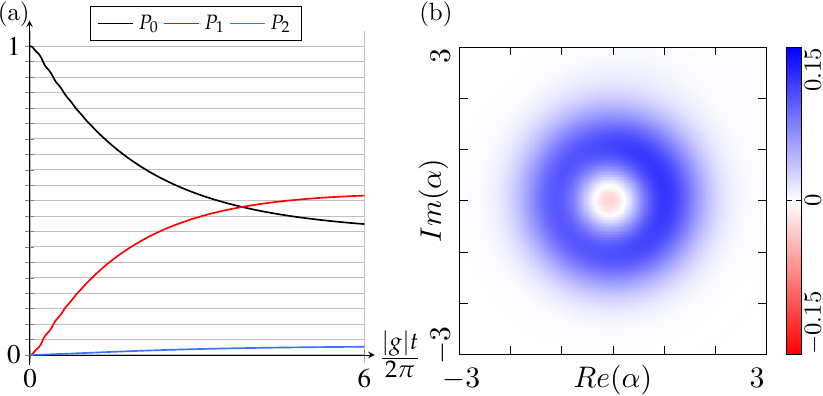}
\caption{(Color online) Analysis of resulting state from the quantum Zeno dynamics for the open system driven at $E^{(2)}/g=0.75$ at frequency $\omega_{L}=\omega+g+\chi$ and $D^{(2)}/g=0.065$ at frequency $\Omega_{D}=\Omega$. a) Probability to find zero (black), one (red), and two (blue) phonons in the mechanical subsystem over time showing a strong suppression of two or more phonons. b) Resulting Wigner function of the final state displaying a clear negative patch signalling the non-Gaussian nature of the state in the mechanical subsystem.}
\label{fig:ZenoStateEXTRA}
\end{figure}

We study the impact of the additional nonlinearity on the Zeno blockade by comparing the simulations outlined in the main text with the results for the same system but an additional nonlinearity characterised by $\chi=0.2$ MHz. \Fref{ZenoDynEXTRA} shows the behaviour of an optical drive with amplitudes $E^{(0)}/g=0$ (dashed), $E^{(1)}/g=0.25$ (dotted), and $E^{(2)}/g=0.75$ (solid) at a frequency of $\omega_{L}=\omega+2g+4\chi$ which establishes a blockade for two phonons. The results for a resonant mechanical drive at frequency $\Omega_{D}=\Omega$ with amplitude $D^{(2)}/g=0.065$ strongly resemble \fref{ZenoDyn} and feature the negativity of the Wigner function at intermediate driving amplitude. 
Similarly, \fref{ZenoStateEXTRA} illustrates that a drive addition of the nonlinearity for the identical mechanical drive and an optical drive that establishes the blockade at for one phonon at frequency $\omega_{L}=\omega+g+\chi$ and amplitude $E^{(2)}/g=0.75$ do not differ significantly from the results without the perturbing nonlinearity in \fref{ZenoState}.

\section{Multitone state preparation}
\label{sec:AppC}
The quantum Zeno blockade for intermediate couplings was found to suppress more than $M$ excitations in the respective subsystem rather than remove the subspace with exactly $M$ excitations which can be attributed to the non-adiabatic correction $\hat{U}_{\text{na}}$ of the total evolution for finite coupling strengths~\cite{Facchi2003}
\begin{align}
\hat{U}_{K}(t)=\hat{U}_{\text{ad},K}(t)+\frac{1}{K}\hat{U}_{\text{na},K}(t),
\end{align}
where the non-adiabatic correction $\hat{U}_{\text{na}}(t)$ is given by
\begin{align}
\hat{U}_{\text{na},K}(t)=\bigg[\sum_n\sum_{k\ne n}\frac{\hat{P}_k\hat{H}_o\hat{P}_n}{\eta_k-\eta_n},\hat{U}_{\text{ad},K}(t)\bigg]+\mathcal{O}(K^{-1})
\end{align}
while the adiabatic evolution $\hat{U}_{\text{ad},K}(t)$ is governed by 
\begin{widetext}
\begin{align}
\hat{U}_{\text{ad},K}(t)=\exp\bigg(-i\bigg(K\hat{H}_{\text{m}}+\sum_n\hat{P}_n\hat{H}_{o}\hat{P}_n+K^{-1}\sum_n\sum_{k\ne n}\frac{\hat{P}_n\hat{H}_o\hat{P}_k\hat{H}_o\hat{P}_n}{\eta_n-\eta_k}+\mathcal{O}(K^{-2})\bigg)t\bigg).
\end{align}
\end{widetext}
However, the simulations in the main text suggest that the transition from a state below $M$ excitations into the subspace with $M$ excitations is suppressed as well as the transition from $M$ excitations to the subspace with $M+1$. Therefore, we find that the transitions from a state with fewer than $M$ to a state with more than $M$ excitations is suppressed twice and therefore more effectively blocked. Thus, establishing multiple blockades that remove $M-1$ and $M$ excitations in the strong coupling limit can improve the effective blockade for $M$ excitations. Therefore, we employ the driving Hamiltonian 
\begin{align}
\hat{H}_{o}(t)=[i(E_1e^{i\omega_1t}+E_2e^{i\omega_2t}) \hat{a}+i\tilde{D}e^{i\Omega_Dt} \hat{b}+\text{h.c.}].
\end{align}
with $\omega_1=\omega+g+\chi$ and $\omega_2=\omega+2g+4\chi$ to establish the combined blockade and $\Omega_D=\Omega$ to resonantly drive the mechanical subsystem. We employ $\omega/2\pi=5$ GHz and decay rate $\kappa/2\pi=6.48$ kHz for the optical mode while the mechanical mode, the coupling constants and mean occupations of the baths remain the same. Using the drive amplitude for the blockade $E_1/g=0.075$ of the first level and $E_2/g=0.497$ of the second level, as well as the drive amplitude of the mechanical system $\tilde{D}/g=0.0497$ results in the dynamics for the probabilities and Wigner function illustrated in \fref{ZenoIdeal}. In comparison to the blockade with a single drive tone, we find that the probability to find two phonons decays beyond the first Rabi-like oscillation towards a improved blockade at the final time $t_F=8\pi/|g|$ with $1-P_0-P_1=0.0216$, as displayed in \fref{ZenoIdeal}(a). Moreover, we find the probability of the Fock states $P_1(t_F)=0.905$ and $P_2(t_F)=0.018$. These results demonstrate that the suggested QZE scheme enables the creation of Fock states for systems which need to improve only the cavity linewidth by one order of magnitude beyond current state of the art. The corresponding Wigner function is presented in \fref{ZenoIdeal}(b) which shows the similarity of the simulated state to the Fock state $\ket{1}$ as it consists of a large negative patch around the origin in phase space enclosed by a positive with an overall rotation symmetry.
\begin{figure}[H]
\centering
\includegraphics[width=0.48\textwidth]{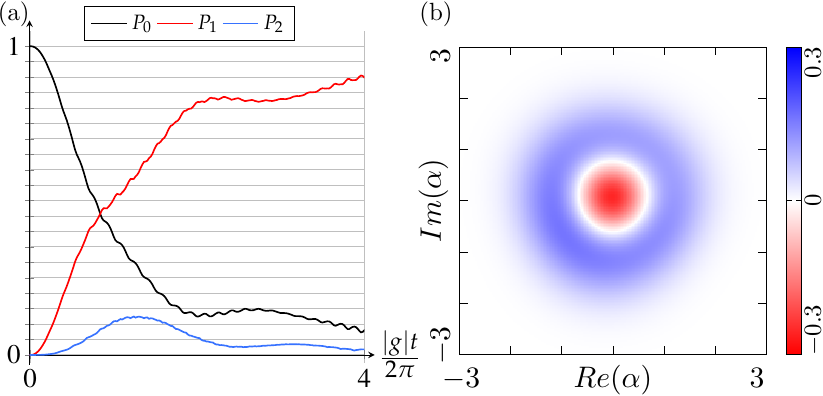}
\caption{(Color online) Analysis of resulting state from the quantum Zeno dynamics with multiple blocking tones in the open system. a) Probabilities for no phonon (black), one phonon (red), and two phonons (blue) over time. b) Resulting Wigner function of the final state.}
\label{fig:ZenoIdeal}
\end{figure}

\end{document}